\documentclass[fleqn,usenatbib]{mnras}

\usepackage{newtxtext,newtxmath}
\usepackage{adjustbox}

\usepackage[T1]{fontenc}
\DeclareRobustCommand{\VAN}[3]{#2}
\let\VANthebibliography\thebibliography
\def\thebibliography{\DeclareRobustCommand{\VAN}[3]{##3}\VANthebibliography}

\usepackage{subcaption}  
\usepackage{afterpage}
\usepackage{caption}     

\usepackage{graphicx}	
\usepackage{amsmath}	
\usepackage{wasysym}
\usepackage{adjustbox}
\usepackage{stfloats}

\title[Infrared Volcanic Satellites]{Volcanic Satellites Tidally Venting Na, K, SO$_2$ in Optical \& Infrared Light}

\author[Oza et al.]{
Apurva V. Oza$^{1,2}$\thanks{E-mail: oza@caltech.edu},
Andrea Gebek$^{3}$,
Moritz Meyer zu Westram$^{4}$, Armen Tokadjian$^{2}$, Anthony L. Piro$^{5}$,   
\newauthor  Renyu Hu$^{2,1}$, Athira Unni$^{6,7}$, Raghav Chari$^{8}$, Aaron Bello-Arufe$^{2}$, Carl A. Schmidt$^{9}$,   Amy J. Louca$^{10}$, 
\newauthor Yamila Miguel$^{10,11}$, Raissa Estrela$^{2}$, Jeehyun Yang$^{2}$, Mario Damiano$^{2}$, Yasuhiro Hasegawa$^{2}$, Luis Welbanks $^{12}$,
\newauthor  Diana Powell$^{13}$, Rishabh Garg$^{14}$, Pulkit Gupta$^{15}$,  Yuk L. Yung$^{1}$, 
   and Rosaly M.C. Lopes$^{2}$
\\
$^{1}$Division of Geological and Planetary Sciences, California Institute of Technology, Pasadena, USA\\
$^{2}$Jet Propulsion Laboratory, California Institute of Technology, Pasadena, USA\\
$^{3}$Sterrenkundig Observatorium, Universiteit Gent, Ghent, Belgium \\
$^{4}$ Physikalisches Institut, Universität Bern, Bern, Switzerland \\
$^{5}$The Observatories of the Carnegie Institution for Science, Pasadena, 91101 \\
$^{6}$Department of Astronomy \& Astrophysics, University of California, Santa Cruz, USA \\
$^{7}$Department of Physics and Astronomy, University of California, Irvine, USA \\
$^{8}$Department of Physics and Astronomy, University of Tennessee, 
Knoxville, USA\\ 
$^{9}$Center for Space Physics, Boston University, Boston, USA \\ 
$^{10}$ Leiden Observatory, Leiden University, Leiden, The Netherlands \\
$^{11}$ SRON Netherlands Institute for Space Research, Leiden, The Netherlands \\
$^{12}$School of Earth and Space Exploration, Arizona State University, Tempe, USA \\ 
$^{13}$Department of Astronomy \& Astrophysics, University of Chicago, Chicago, USA \\
$^{14}$Portola High School, Irvine, USA \\
$^{15}$Indian Institute of Technology, Indore, India \\
}

\date{Accepted 4 August 2025, in press}
\pubyear{2025}

\newcommand{\psj}{PSJ}
\begin{document}
\label{firstpage}
\pagerange{\pageref{firstpage}--\pageref{lastpage}}
\maketitle

\begin{abstract} 
Recent infrared spectroscopy from the \textit{James Webb Space Telescope} (JWST) has spurred analyses of common volcanic gases such as carbon dioxide (CO$_2$), sulfur dioxide (SO$_2$),  alongside alkali metals sodium (Na I) and potassium (K I) surrounding the hot Saturn WASP-39 b. We report more than an order-of-magnitude of variability in the density of neutral Na, K, and SO$_2$ between ground-based measurements and JWST, at distinct epochs, hinting at exogenic physical processes similar to those sourcing Io's extended atmosphere and torus. Tidally-heated volcanic satellite simulations sputtering gas into a cloud or toroid orbiting the planet, are able to reproduce the probed line-of-sight column density variations. The estimated SO$_2$ flux is consistent with tidal gravitation predictions, with a Na/SO$_2$ ratio far smaller than Io's. Although stable satellite orbits at this system are known to be $<$ 15.3 hours, several high-resolution alkali Doppler shift observations are required to constrain a putative orbit. Due to the Roche limit interior to the planetary photosphere at $\sim$8 hours, atmosphere-exosphere interactions are expected to be especially important at this system. 
\end{abstract}


\section{Introduction} \label{sec:intro}

At the time of this writing, Saturn has 274 confirmed moons in orbit, 67 of which were discovered in the past decade alone \citep{Sheppard2018, Sheppard2023}, and 128 more were reported by the Minor Planet Center on March 11, 2025 following an irregular satellite campaign at the CFHT \citep{Ashton2025}. It is now well known that Saturn's rings are actively shaped by its natural satellites \citep{charnoz18book}. Another splendid example is Enceladus, which actively shapes Saturn's E-ring due to its tidally-heated hydrothermal venting, leading to a hydroxide torus of gas \citep{johnson06a} and grains. Of course, Io is well known to host a $\mu$bar volcanic and sublimated sulfur dioxide SO$_2$ atmosphere observed to vary from 10$^{16}$ - 10$^{17}$ SO$_2$/cm$^2$\citep{lellouch15}. In fact, SO$_2$ is one of the most common gases, along with CO$_2$ and H$_2$O, released in a volcanic eruption on Earth. Ongoing volcanism on Venus was only recently confirmed \citep{Herrick2023, Sulcanese2024} despite its obvious atmospheric SO$_2$ signature with \textit{Venus express} measurements at 0.02 ppm below 90 km and roughly 5 ppm above 100 km \citep{Mahieux2023}. 
In terms of discovered elements at exoplanets, the relative abundances of volatile metals, notably neutral sodium (Na I), potassium (K I) and SO$_2$ are well known to be sensitive to silicate lava temperatures \citep{fegleyzolotov2000}, reminding us of one of their origins as silicate magmas.

\begin{figure*} 
    \centering
    \includegraphics[width=\textwidth, height=6.0cm]{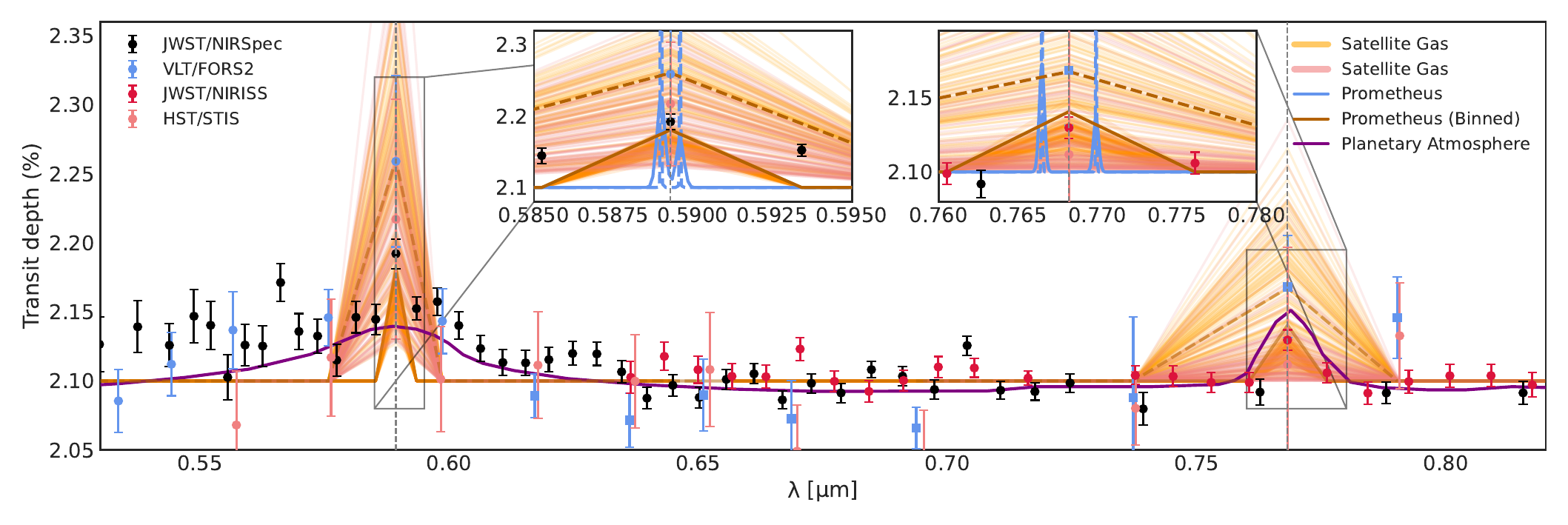}
    \caption{Alkali metals: Na I \& K I as observed by the VLT (blue) \citep{Nikolov2016}, HST (pink) \citep{Fischer2016},  JWST/NIRISS (red) \citep{Feinstein23}, and JWST/NIRSpec (black) \citep{Rustamkulov23}, with accompanying planetary (1-D radiative-convective-thermochemical equilibrium: purple) and satellite gas (orange for VLT and JWST, pink for HST) simulated with 100 optically-thin models c.f. Eqn \ref{eq:opticallyThinR} approximating equivalent widths W$_{\lambda} \propto N$ (Eqn. \ref{equationW}) reported in Table \ref{tab:obs_table}. Zoom inset: \texttt{prometheus} forward models are shown in high-resolution (satellite cloud: blue-dashed; torus: blue solid) and binned to the data (brown in the respective line styles). }
    \label{fig0:NaK}
\end{figure*}
In principle, if a third body were to be present at a planetary system, the natural satellite would experience tidal heating should it be in a perturbed or eccentric orbit \citep{peale79}.


\begin{table*}
	\centering
    \begin{adjustbox}{width=\linewidth}
	\label{tab:posterior_framework}
    \setlength{\tabcolsep}{5pt}
	
\begin{tabular}{lccccccc}
		\hline\hline
		Telescope (Alkali) & VLT (Na) & HST (Na) & JWST (Na) & VLT (K) & HST (K) & JWST (K) & JWST (K) \\
          Instrument   & FORS2 & STIS & NIRSpec & FORS2 & STIS & NIRISS & NIRSpec \\
		\hline
        Epoch & 2016-3-8 & 2013-3-17 & 2022-7-10 & 2016-3-12 & 2013-3-17 & 2022-7-27 & 2022-7-10 \\
        $dF_{\lambda}/F_{\star}$ (\%) & $2.234^{+0.067}_{-0.067}$ & $2.189^{+0.078}_{-0.050}$ & $2.190^{+0.011}_{-0.011}$ & $2.143^{+0.042}_{-0.039}$ & $2.095^{+0.040}_{-0.021}$ & $2.127^{+0.008}_{-0.008}$ & $2.125^{+0.001}_{-0.001}$ \\
        $\log N$ [cm$^{-2}$]         \textbf{MCMC} & $11.76^{+0.18}_{-0.28}$ & $11.54^{+0.27}_{-0.33}$ & $10.96^{+0.43}_{-0.10}$ & $11.46^{+0.26}_{-0.61}$ & $10.94^{+0.62}_{-0.63}$ & $10.62^{+0.19}_{-0.17}$ & $9.96^{+1.36}_{-1.36}$ \\
		$\sigma_{v}$ [km/s]& $31.6^{+116.3}_{-25.2}$ & $32.4^{+119.0}_{-25.8}$ & $89.1^{+124.7}_{-50.2}$ & $28.8^{+112.4}_{-22.5}$ & $30.2^{+117.7}_{-23.9}$ & $11.5^{+39.8}_{-9.1}$ & $0.98^{+1.16}_{-0.52}$ \\
		$v_{kin}$ [km/s] & 50.5 & 51.6 & 142.2 & 46.0 & 48.2 & 18.3 & 1.56 \\
        \hline
        $\log{N}$ [cm$^{-2}]$  \textbf{\texttt{prometheus}} & 11.45 ($\rightmoon$) & - & 10.77 (torus) & 11 ($\rightmoon$) & - & 10.38 (torus) & - \\
        $\sigma _v$ [km/s] & 7.8 & - & 62.7 & 7.8 & - & 62.7 & - \\
        $v_{kin}$ [km/s] & 12.5 & - & 100 & 12.5 & - & 100 & - \\
		\hline\hline
	\end{tabular}
    \end{adjustbox}
    \caption{Optically-thin alkali column density minima $N$, corresponding 1-D velocity dispersions $\sigma_v$, and approximate kinetic velocities (Eqn. \ref{eqn:vMB}) for \texttt{MCMC} posterior simulations (top; (orange in Figure \ref{fig0:NaK})) and \texttt{prometheus} models (bottom; blue, brown). Transit depths are normalized at $c = 0.0210$. Satellite cloud and tori geometries are indicated as ($\rightmoon$) and (torus) respectively. }
    \label{tab:alkalis}
\end{table*}

Especially exciting is the case for a habitable-zone exomoon \citep{TokadjianPiro2023}, possibly outshining the planet itself from 6-8 $\mu$m in the infrared, due to the satellite blackbody flux being boosted by tidal heating.

Novel techniques to detect infrared-bright exomoons in spectroscopy  \citep[e.g.,][]{Elina2023, Elina2024}, has spawned multiwavelength searches in radio \citep{narang2023a, narang2023b} and thanks to the bright sodium and potassium lines in absorption spectroscopy, visible light \citep{oza2019b}. At transiting exoplanet systems, putative exomoons are subject not only to the host planet's gravitational tide, but also the massive gravitational field of the host star. Beyond a planetary orbit of a couple days, tides dissipate energy into the interior of the satellite due to orbital resonance, rather than decay its orbit (\citep{cassidy09}, see also Figure \ref{fig:moonstable}). The tidal field of the star and planet in concert, acting on a natural satellite, drive three-body gravitational heating, an expression derived in \citet{cassidy09, oza2019b} which is far larger at a close-in system than canonical eccentricity-driven tides \citep{peale79}. 3-body heating is so large that gas and grain volcanism on these putative satellites is inevitable and predicted to be roughly $\sim$ 10$^{5 \pm 1} \times$ the energy measured at Io \citep{Veeder1994}. As a first approximation, if the satellite has an unknown size $R_{\rightmoon} \sim R_{Io}$, orbiting a Jupiter or Saturn-sized object over a range of orbital periods $ 2 \lesssim \tau_{p} \lesssim 5$ days (equilibrium temperatures of $\sim 1000 \lesssim T_{eq} \lesssim 2500$ Kelvins), the tidal venting rate computed by \citet{oza2019b} is roughly $\approx$ 10$^{8 \pm 1}$ kg/s (more than 100 $\times$ Io's). Of course, depending on the size and magmatic state of putative rocky exomoons, thermal evaporation may also source $\dot{M} \gg$ 10$^{8}$ kg/s \citep{oza2019b}. Whether this range of volcanism is observable, and whether any of these putative exomoons have survived disintegration, leads us to the current investigation of infrared data with \textit{James Webb Space Telescope} (JWST) MIR and NIR data.

We focus on one of the farthest host planets from the sample of extrasolar-Ios studied by \citet{oza2019b} at $\tau_p \sim$ 4.1 days. The distance allows for a larger Hill sphere where natural satellites are dynamically stable \citep{Kisare2024}. WASP-39 b is a highly-inflated hot Saturn orbiting a m$_K$=12.11, G7 host star located 215.4 $\pm$ 0.7 pc away \citep{Faedi2011}. Fortunately WASP-39 b has been heavily monitored in visible light (by the Hubble Space Telescope (HST/STIS) \citep{Sing2016, Fischer2016} VLT/FORS2 \citep{Nikolov2016}), and by JWST in the near-infrared (NIR) \citep{Wakeford2018, Rustamkulov23, Feinstein23, Ahrer2023, Alderson2023}, and mid-infrared (MIR) \citep{Powell2024} for 11 total epochs. In this short letter, we provide a volcanic alternative to the robust H$_2$S driven atmospheric photochemistry \citet{Yung1999, Tsai2023, YangAndHu2024} as a source for SO$_2$. In this light, we determine the optically-thin Na I, CO$_2$, and SO$_2$ column densities from available data using the technique of evaporative transmission spectroscopy  \citep{go2020} in \S \ref{sec:alkali_analysis}. 
Lastly, as a test case in \S \ref{sec:serpens} we simulate the toroidal exosphere of WASP-39 b sourced by an outgassing rate $\dot{M}$ via the exomoon torus softwares: \texttt{dishoom} \& \texttt{serpens} \citep{MzW24}. Finally, we discuss implications and future outlook in \S \ref{sec:discussion}. 

\begin{table*}
    \centering
    \begin{tabular}{lccccc} 
        \hline
        Epoch & Atom & Instrument & N [cm$^{-2}$] & \textbf{W$_{\lambda}$} (mÅ) & $\dot{M}_{\mathrm{tor}} - \dot{M}_{\gamma}$ [kg/s] \\
        \hline
        2016-03-08 & Na & VLT/FORS2 & $10^{11.65^{+0.19}_{-0.36}}$ & $131 \pm 73$ & $(0.8 - 81) \times 10^{5}$ \\
        2013-03-17 & Na & HST/STIS/G750L & $10^{11.60^{+0.27}_{-0.81}}$ & $119 \pm 109$ & $(0.3 - 87) \times 10^{5}$ \\
        2022-07-10 & Na & JWST/NIRSpec & $10^{10.78^{+0.11}_{-0.15}}$ & $17.8 \pm 5.2$ & $(0.2 - 9.1) \times 10^{4}$ \\
        2016-03-12 & K  & VLT/FORS2 & $10^{11.42^{+0.29}_{-1.43}}$ & $136 \pm 131$ & $(0.7 - 380) \times 10^{5}$ \\
        2013-03-17 & K  & HST/STIS/G750L & $< 10^{10.43}$ & $14.0 \pm 233.6$ & $< 35 \times 10^{5}$ \\
        2022-07-10 & K  & JWST/NIRSpec & $< 10^{10.47}$ & $15.5 \pm 28.7$ & $< 6.2 \times 10^{5}$ \\
        2022-07-27 & K  & JWST/NIRISS & $10^{10.61^{+0.12}_{-0.17}}$ & $21.3 \pm 7.0$ & $(0.2 - 39) \times 10^{5}$ \\
        \hline
        Epoch & Molecule & Instrument & N [cm$^{-2}$] & T$_{\mathrm{kin}}$ (K) & $\dot{M}_{\mathrm{tor}} - \dot{M}_{\gamma}$ [kg/s] \\
        \hline
        2022-07-30 & SO$_2$ & JWST/G395H & $10^{16.7^{+0.1}_{-0.1}}$ & $1100^{+120}_{-110}$ & $(0.6 - 40) \times 10^{11}$ \\
        2022-07-10 & SO$_2$ & JWST/NIRSpec & $10^{16.4^{+0.2}_{-0.5}}$ & $780^{+460}_{-330}$ & $(0.3 - 20) \times 10^{11}$ \\
        2023-02-14 & SO$_2$ & JWST/MIRI & $10^{15.1^{+0.2}_{-0.5}}$ & $1180^{+1400}_{-690}$ & $(0.15 - 10) \times 10^{10}$ \\
        2022-07-30 & CO$_2$ & JWST/G395H  & $10^{15.46^{+0.03}_{-0.03}}$ & $2200^{+140}_{-100}$ & $(2.4 - 3.0) \times 10^{8}$ \\
        2022-07-10 & CO$_2$ & JWST/NIRSpec  & $10^{15.44^{+0.02}_{-0.02}}$ & $2000^{+70}_{-90}$ & $(2.3 - 2.9) \times 10^{8}$ \\
        \hline
    \end{tabular}
    \caption{Optically thin column densities for various volcanic species. For atoms we report the equivalent width (W$_\lambda$) in mÅ, and for molecules we report the temperature $T_{\mathrm{kin}}$ in Kelvins. NIR retrievals are from 3.9--4.5\,$\mu$m, MIR retrievals from 6--8\,$\mu$m. Normalizations: JWST/PRISM $2.107^{+0.006}_{-0.007}$, JWST/G395H $2.089^{+0.006}_{-0.006}$, JWST/MIRI $2.108^{+0.007}_{-0.007}$. Column densities $N$, are estimated by transit depths as described for atoms via W$_{\lambda}$ in \S \ref{sec:alkaligas} and inverse models for molecules, in \S \ref{sec:volcanicgas}. $\dot{M}$ represents the inferred mass loss rates. $\gamma$- and torus-limited describe photoionization-limited clouds\footref{foot:cet} and tori with $\sim$3h charge-exchange lifetimes.}
    \label{tab:obs_table}
\end{table*}

\vspace{-2.5em} 
\section{Evaporative Transmission Spectra Analysis}\label{sec:alkali_analysis}
We first consider simplistic optically thin absorption from atoms and molecules in a  tenuous exosphere. In this optically-thin scenario, the transit depth is directly proportional to the number of absorbing particles spread over the stellar disk, expressed as the change in stellar flux over a wavelength region $\lambda$ \citep{draine2011, go2020}:
\begin{equation}\label{eq:opticallyThinR}
    \frac{dF_{\lambda}} {F_{\star}} \approx N \sigma(\lambda),
\end{equation}

where $\sigma(\lambda)$ is the absorption cross-section, and $N$ (cm$^{-2}$) the average line-of-sight (LOS) column density of the absorber. Figure \ref{fig0:NaK} shows how the transit depth $\frac{dF_{\lambda}} {F_{\star}}$ near the alkali absorption lines varies between epochs (black, blue, and red data points). 

As described more in Section \ref{sec:serpens} the total number of absorbing particles can be estimated by a mass loss relation: 

\begin{equation}\label{eq:NNmdot}
    \mathcal{N} = (\dot{M}/m) \cdot \tau_\gamma
\end{equation}

where $m$ is the volatile mass, $\tau_\gamma$ is the lifetime against photoionization, approximating a \textit{cloud} of particles limited by the ambient solar field. More generally, $\tau_\gamma$ can be described for an arbitrary species $\tau_i$ and physical processes\footnote{$\tau_{\gamma, CO_2} = 2.3h; \tau_{\gamma, SO_2} = 2.5m ; \tau_{\gamma, Na} = 6.6m ; \tau_{\gamma, K}$=1.8m \label{foot:cet}} \citep{oza2019b}. To approximate a \textit{toroid} we adopt the value $\tau_{tor} \sim$ 3 hours by \citet{MeyerZuWestram2023} characteristic of charge-exchange \citep{JohnsonStrobel1982} and other well-known processes at Jupiter-Io plasma torus \citep{schmidt23}. In this way $\dot{M}_{\gamma}$ and $\dot{M}_{tor}$ are inferred mass loss rates to power the observed line-of-sight column densities $\bar{N} \sim \frac{\mathcal{N}}{\pi R_\star ^2}$ so that the column density is directly proportional to the mass loss rate:

\begin{equation}\label{eq:column_massloss}
      N \sim \dot{M} \frac{\tau_i }{m_i \pi R_{\star}^2}
\end{equation}

where $\mathcal{N}$ (Eqn. \ref{eq:NNmdot}) can be distributed into a photoionized cloud or plasma torus geometry \citep{go2020}. 

Table \ref{tab:obs_table} shows (a) for atomic features, the computed column densities (Eqn. \ref{eq:opticallyThinR}) for each epoch of data and equivalent widths (see Section \ref{sec:alkaligas}); (b) for molecular features, column densities and temperatures (Section \ref{sec:volcanicgas}), and (c) mass loss rates for both atoms and molecules (as described by Eqns. \ref{eq:NNmdot} and \ref{eq:column_massloss}).
 
\vspace{-1.5em} 
\subsection{Alkali Gas in the Optical}\label{sec:alkaligas}
One can approximate $N$ of neutral sodium and potassium analytically by the sum of two Voigt profiles (approximating the D$_2$ and D$_1$ alkali resonance lines). As the low-resolution data here does not resolve these absorption lines, we measured the equivalent width (W$_{\lambda}$) of the line relative to the continuum determined by the normalization fit to the data near the Na \& K resonance lines. W$_{\lambda}$  therefore is directly proportional to N, given by \citet{draine2011}

\begin{equation}\label{equationW}
    W_\lambda=\frac{\pi e^2}{m_ec^2}(f_\mathrm{D2}\lambda_\mathrm{D2}^2+f_\mathrm{D1}\lambda_\mathrm{D1}^2)\times N,
\end{equation}
where $e$ the elementary charge, $m_e$ the electron mass, $c$ the speed of light, and $f_\mathrm{D}$ ($\lambda_\mathrm{D}$) the oscillator strength (wavelength) of the D-line alkali transition, as also summarized in \citet{oza2019b} Eqn. 14. This is an approximation as discussed by \citet{draine2011} is exact for $\tau_0 \rightarrow$ 0, and within 2.6 \% for the optically thin $\tau_0 < $1.256, in the limit where the term $\left ( \frac{1}{1 + \frac{\tau_0} {2\sqrt(2)}} \right ) $ approaches unity.  It follows that the measured equivalent width of the sodium and potassium lines are directly related to the average line-of-sight column density $N$ reducing to the number of absorbing atoms $\mathcal{N} \sim N \pi R_{\star}^2 $ \citep{johnsonhuggins06} modeled in 3-D in Section \ref{sec:serpens}. For each individual transit of WASP-39 b, neutral sodium \& potassium column densities are simulated with 100 optically-thin models following Equation \ref{eq:opticallyThinR} indicated in orange and pink corresponding to the datasets identified in the caption, approximating the range of column densities derived from their equivalent widths following Equation \ref{equationW} \citep{draine2011}, reported in Table \ref{tab:obs_table}. \texttt{Prometheus} simulations \citep{go2020} (indicated in blue, brown (binned)) approximate gas density minima values provided in Table \ref{tab:alkalis}. 

\par In Figure \ref{fig0:NaK} we employ the radiative transfer tool \texttt{prometheus} to model the transit depth of a satellite gas source following \citet{go2020}, distributing a packet of evaporating $\mathcal{N}$ particles into two distinct geometries: 1) toroid (solid blue) and 2) sputtered cloud (dashed blue). Due to the tenuous nature and limited extent of the cloud scenario, the dashed absorption lines are narrow probing mostly the line core.
\begin{figure*}
    \centering
    \includegraphics[width=\linewidth]{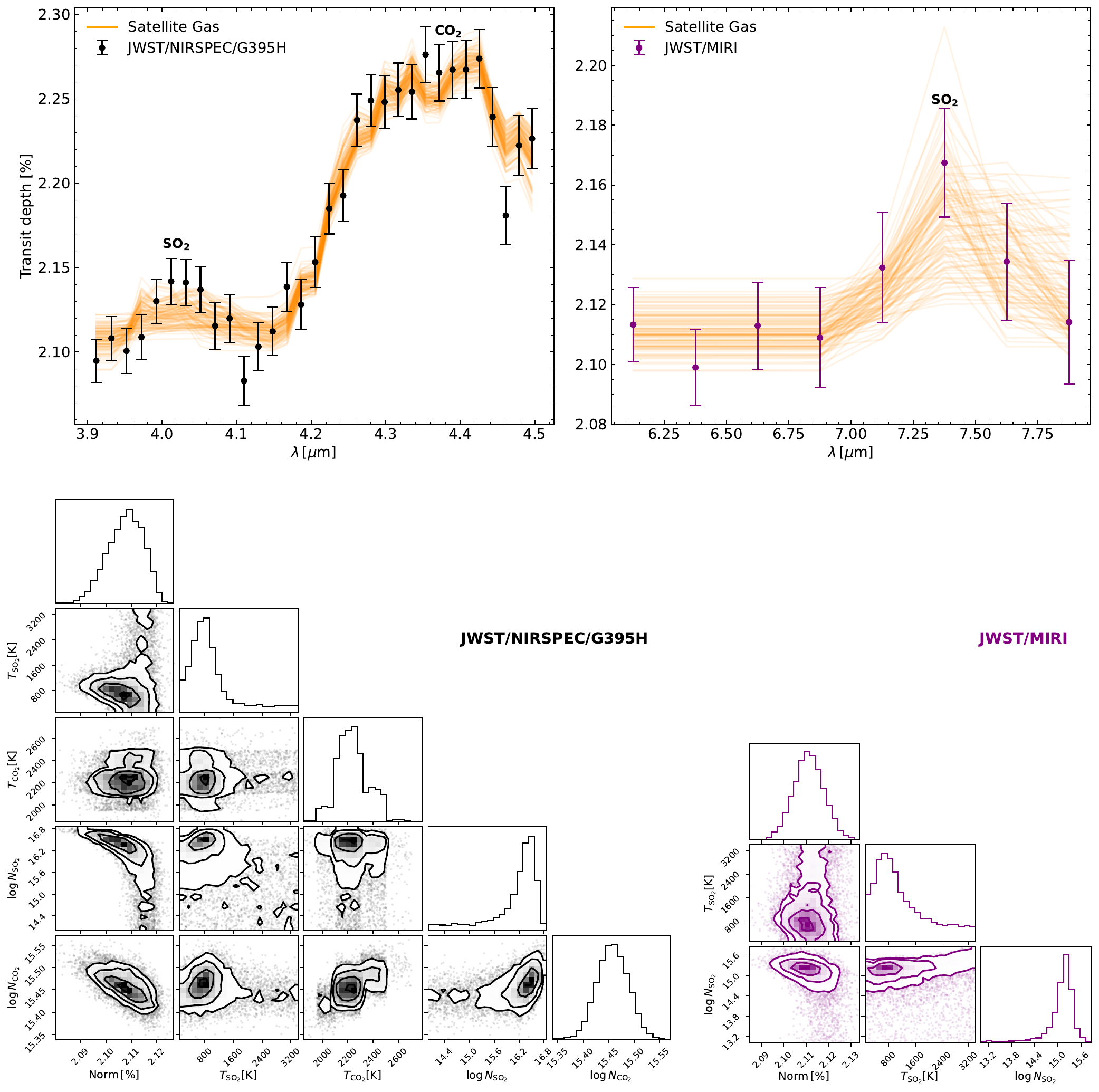}
    \caption{\textit{Top panel:} 100 optically-thin transmission spectra models of SO$_2$ and CO$_2$ gas in orange, with central values listed in Table \ref{tab:obs_table}.\textit{ Bottom panel:} Posterior distributions, including the temperature T of the gas, column density N, and normalization. JWST NIR data is in black and MIR in purple.}
    \label{fig1:optically-thin}
\end{figure*}
The planetary atmosphere (purple) is adopted from \citet{Rustamkulov23} assuming chemical equilibrium in a hydrostatic atmosphere.  The simulated minimum alkali column densities are computed here from Eqn. \ref{equationW} (Table \ref{tab:obs_table}) based on the sodium and potassium data seen by VLT, HST, and JWST. Strikingly the data are not consistent within error. Moreover, we find that the equivalent width changes by a factor of 7.4 from epoch to epoch for sodium, and more than 1.3 to a surprising disappearance for potassium.  Two \texttt{prometheus} models are shown in Figure \ref{fig0:NaK}: 1) an alkali toroid (solid blue) with a semi-major axis, a$_{torus} \sim$ 1.5 R$_p$ and 1-D velocity dispersion $\sigma_v \sim$ 62.7 $\mathrm{km\,\mathrm{s}^{-1}}$, which if due to thermal motions, described in \citet{go2020} Eqn. 2 as a line temperature,  follows a Maxwell Boltzmann distribution with average velocity: $\bar{v}_{MB} = \sqrt{\frac{8k_bT}{\pi m}}$ \citep{reif1965} so that:

\begin{equation}\label{eqn:vMB}
  v_{kin} = \sqrt{\frac{8}{\pi}} \, \sigma_v  
\end{equation}

roughly 100 km/s, characteristic of energetic neutral atoms (ENAs) and plasma interactions at Jupiter-Io \citep{MeyerZuWestram2023, wilson02}.  Secondly, 2) an alkali cloud (dashed blue) orbiting at 0.41 R$_{Hill}$ \citep{Kisare2024} fueled by an exomoon at $a_{\rightmoon}$ $\sim$ 1.5 R$_p$ on a $\tau_{\rightmoon} \approx$ 15.3 hour orbit c.f. Table \ref{tab:alkalis}.

Specifically, a satellite cloud (dashed orange \texttt{prometheus} line in Figure \ref{fig0:NaK}) with a column density of $\sim$10$^{11.45}$ Na/cm$^2$ roughly approximates the observed VLT (blue) and HST (pink) absorption values: ($\sim 10^{11.65}$ and $\sim 10^{11.60} \, \mathrm{cm}^{-2}$, respectively), whereas the JWST/NIRSpec (black) epoch is slightly more tenuous $\sim 10^{10.78} \mathrm{cm}^{-2}$, roughly consistent with a torus (solid, orange \texttt{prometheus} line in Figure \ref{fig0:NaK}) fueled by $\sim$ 7.2 $\times$ 10$^{32}$ Na I particles. In the JWST/NIRISS dataset (red), the potassium column is $\sim 10^{10.61} \mathrm{cm}^{-2}$, which \texttt{prometheus} simulations approximate with $\sim$ 3 $\times$ 10$^{32}$ K I particles in a toroidal geometry. Although high-resolution observations are needed to quantify Na \& K variability distinct from a super-stellar planetary atmosphere \citep{Powell2024}, the variability in K I is particularly stark, with detections in the VLT and JWST/NIRISS epochs ($\sim 10^{11.4} - 10^{10.6} \, \mathrm{cm}^{-2}$) contrasted by a a disappearance in the JWST/NIRSpec and HST epochs (Table \ref{tab:obs_table}), which may be suggestive of an orbiting satellite cloud $\sim$ 10$^{12.1}$ $\mathrm{cm}^{-2}$, varying its phase along the observer's line-of-sight \citep{MeyerZuWestram2023}.
\vspace{-1.7em} 

\begin{figure*}
    \centering
    \begin{subfigure}[b]{0.75\textwidth}
        \adjustbox{width=\textwidth, trim=0 0 {0.5\width} 0, clip}{%
            \includegraphics{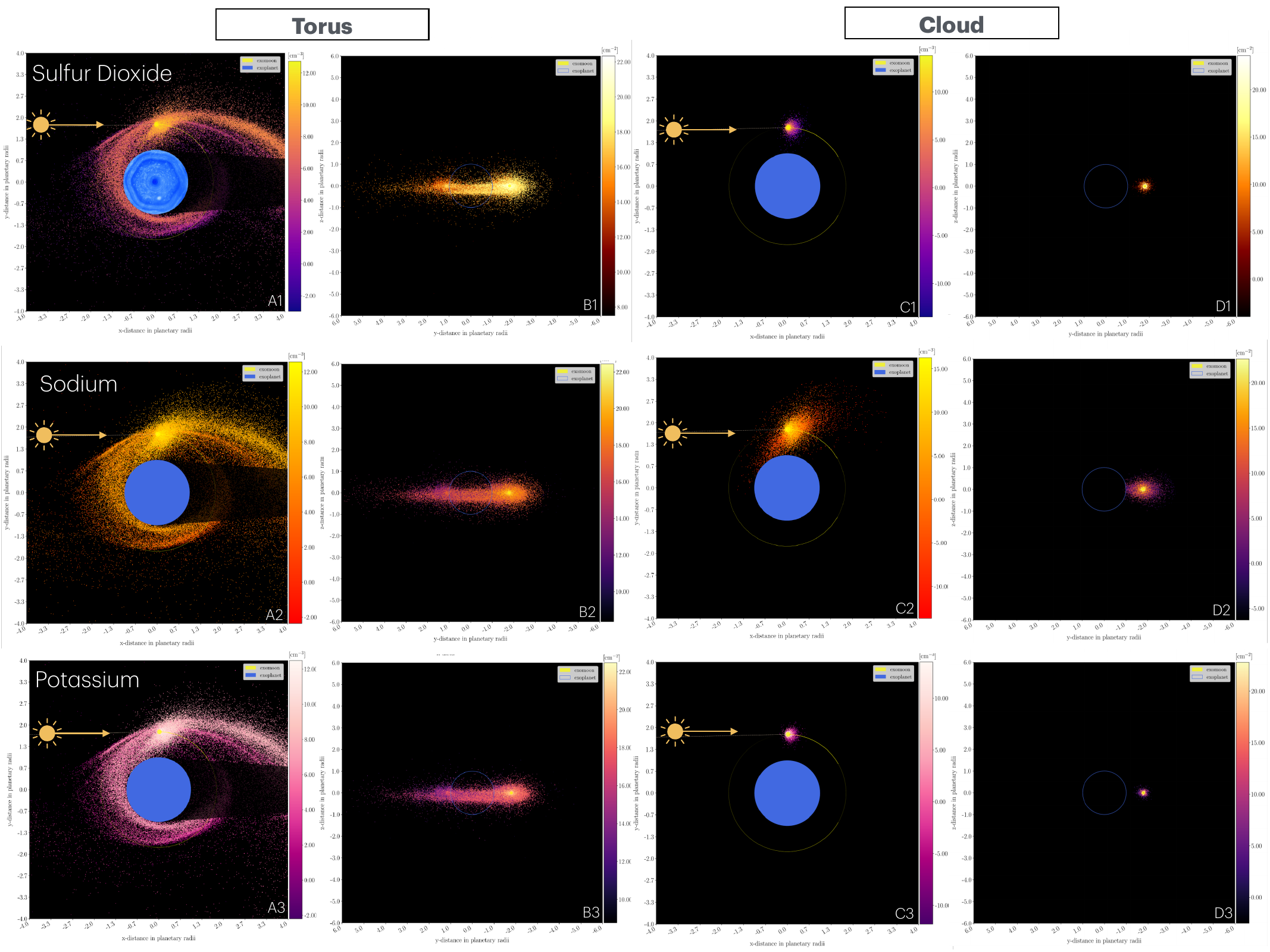}%
        }
        \caption{} 
        \label{fig:serpens_left}
    \end{subfigure}
    
    \caption{
        (\subref{fig:serpens_left}) 
         \texttt{serpens} Monte-Carlo simulations of neutral SO$_2$, Na, K  around an \textit{exo-Io} for a toroidal geometry. Top down (1st column: A1, A2, A3) and line-of-sight (2nd column: B1, B2, and B3) simulations.  
        The approximate total number of particles reproducing the evaporative transmission spectra simulated by \texttt{prometheus} in Section \ref{sec:alkali_analysis} are: $\mathcal{N}_{\mathrm{Na}}= 10^{32.82}$, $\mathcal{N}_{\mathrm{K}}= 10^{32.97}$, and $\mathcal{N}_{\mathrm{SO_2}}= 10^{39}$ atoms/molecules in the system. Particles inside the exoplanet shadow have a lifetime of 3 hours for all cases. }
    \label{fig:serpensNaKSO2}
\end{figure*}

\begin{figure*}
    \ContinuedFloat 
    \centering
    \begin{subfigure}[b]{0.8\textwidth}
        \adjustbox{width=\textwidth, trim={0.5\width} 0 0 0, clip}{%
            \includegraphics{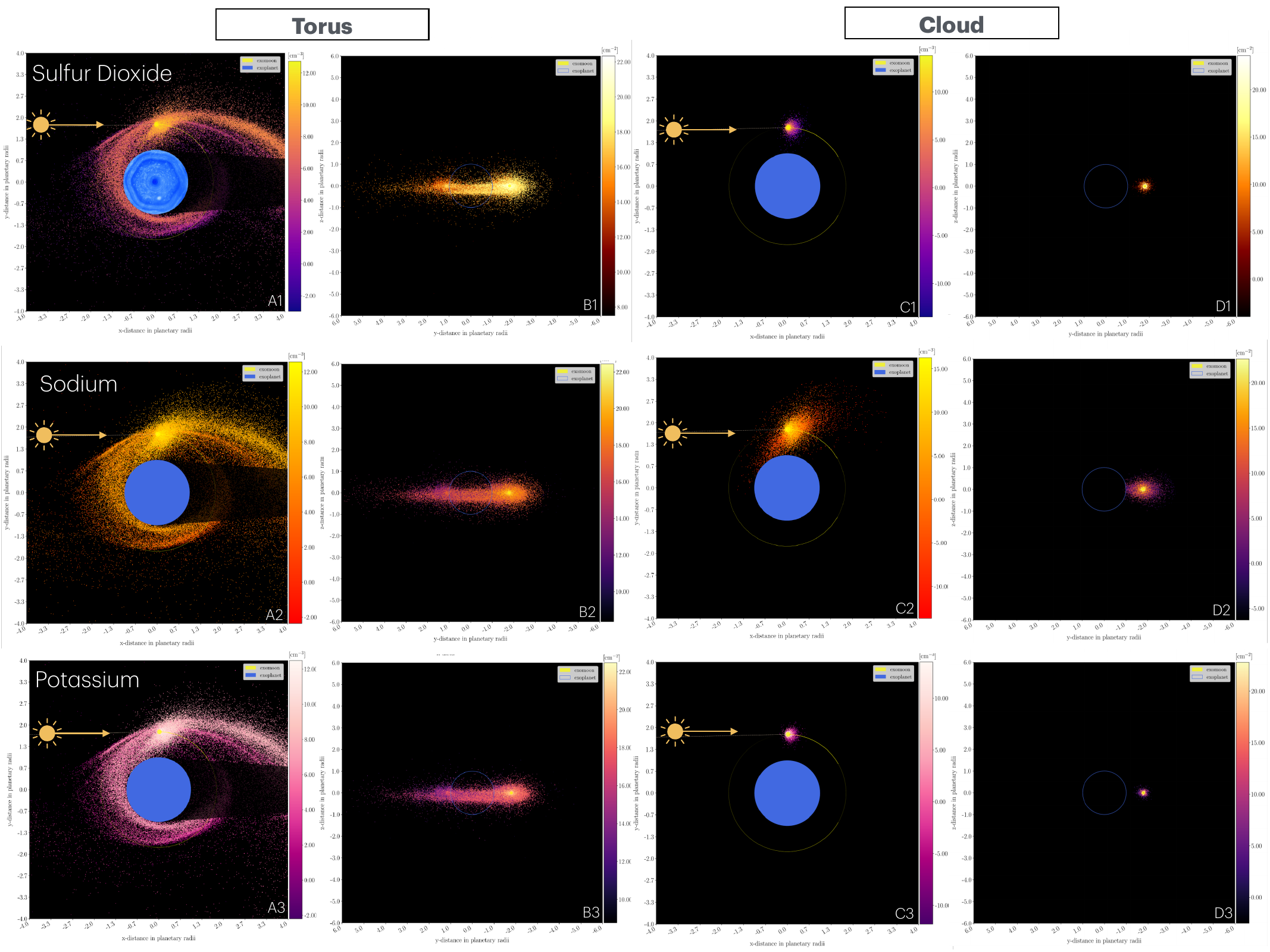}%
        }
        \caption{Cloud geometry. Top down (1st column: C1, C2, C3) and line-of-sight: (2nd column: D1, D2, and D3) simulations.  Species lifetimes in the cloud scenario are limited by photoionization following \citet{huebnermujherjee, oza2019b}.   
        The exoplanet transit duration is given by $t_{1-4} =$ 2.8032 $\pm$ 0.0192 hours \citep{Faedi2011}. 1-D \texttt{serpens} cuts against observations representing minimum optically-thin column densities are described in Figure \ref{fig:panel1} (Na) ,\ref{fig:panel2} (K), and \ref{fig:panel3} (SO$_2$).} 
        \label{fig:serpens_right}
    \end{subfigure}
\end{figure*}
\subsection{Volcanic Gas in the Infrared}\label{sec:volcanicgas}
\vspace{-0.7em} 
For molecular absorption in the NIR and MIR, the approximation in Eqn. \ref{equationW} is more complex, as $\sigma_\lambda$ is no longer an analytical function of wavelength, but rather a composite of many line transitions. We therefore model the observed SO$_2$ and CO$_2$ NIR and MIR transit depth $\frac{dF_{\lambda}} {F_{\star}}$ in the optically-thin limit (Eqn. \ref{eq:opticallyThinR}) by computing $\sigma (\lambda)$ on a T-P-grid from the ExoMOL database \citep{Tennyson2024} in order to infer $N$ and its error. Assigning a temperature and a pressure to the gas is a significant simplification of the problem as the upper atmosphere or exosphere is generally not in local thermodynamical equilibrium (LTE), well known for Io's mid-infrared SO$_2$ bands \citep{lellouch15}. 
Here, we assume the lowest tabulated pressure value of $0.1\,\mathrm{nbar}$ when extracting ExoMOL opacities. We find that increasing pressure up to $0.1\,\mathrm{\mu bar}$ hardly affects the molecular opacities considered here, and are therefore representative of minima.  We leave temperature as a free parameter from $T$ = 100-3400\,K and allow each absorbing species to have a different $T$.  Additionally, we assume a constant normalization, treated as a free parameter. Overall, this leaves us with two free parameters for each absorber contributing to the observed transit depth, plus one free parameter regulating the normalization. We use flat priors for $T$,  $\log_{10}(N/\mathrm{cm}^{-2})$, and $\log_{10}(\sigma_\mathrm{v}/(\mathrm{km\,\mathrm{s}^{-1}}))$. We run a \texttt{python} MCMC \texttt{emcee} sampler \citep{Foreman-Mackey2013} with 32 walkers for 5000 steps, and discard the first 1000 steps as burn-in. The results (column densities and temperatures) from the optically-thin retrieval for the various infrared datasets are summarized in Table ~\ref{tab:obs_table} for molecules, with the accompanying spectra and posterior distributions in Figure \ref{fig1:optically-thin}. 
Whereas an extended atmosphere of WASP-39 b is expected to return a consistent signal across planetary transit epochs, an exomoon orbit naturally leads to order-of-magnitude changes in column density and, consequently, spectral transit depth. For instance the SO$_2$ flux from the 2022-07-30 epoch to 2023-02-14 has varied from N$_{SO_2}$ $\sim$ 10$^{16.7 \pm 0.1}$ to N$_{SO_2}$ $\sim$ 10$^{15.1^{0.2}_{-0.5}}$ SO$_2$/cm$^2$,  far larger than the CO$_2$ variability. Indeed, the CO$_2$ appears to be consistent despite the optically-thin approximation on average near $\sim$ 10$^{15.45 \pm 0.03}$ CO$_2$/cm$^2$ implying the SO$_2$ is indeed suggestive.
Although, WASP-39 has not been reported to be active \citep{Faedi2011}, 2/14 exomoon candidates from the \citet{oza2019b} sample, highlight sources of variability worth considering. Star spot crossings at WASP-52 were shown to modulate H$_2$O abundances \citep{Fournier-Tondreau_2025}, while at WASP-31 b puzzling detections and non-detections of K I by the VLT in both low and high-resolution spectroscopy were reported to be instrumental, albeit with no systematic identified \citep{gibson19}. Multi-epoch observations with the same instrument are therefore essential to characterize variability. Indeed at the volcanic exoplanet candidate, 55 Cancri-e \citep{Quick2020}, toroidal behavior near 4 microns is seen across multiple epochs with the same JWST instrument \citep{Patel2024}.
We now use the optically thin analytic and \texttt{prometheus} forward models as a benchmark for detailed Na, K, and SO$_2$ 3-D Monte Carlo simulations of the approximate toroidal gas geometry in the next section.

\section{Numerical Estimates of Volcanic Gas Densities}\label{sec:serpens}
In addition to fitting the NIR and MIR observations as described in Section \ref{sec:alkali_analysis}, we employ the test-particle Monte-Carlo code \texttt{serpens} \citep{MeyerZuWestram2023, MzW24} for first-order simulation analysis in Figure \ref{fig:serpensNaKSO2}.
We infer Na, K, and SO$_2$ particle- and line-of-sight densities at regular time intervals from the WASP-39 b system which is simulated with the inclusion of a putative exomoon WASP-39 b I as the particle source.
We distinguish between a cloud and toroidal scenario as described in \S \ref{sec:alkali_analysis} on evaporative transmission spectroscopy analysis, where the

density computations are normalized by maintaining a constant total number of atoms $\mathcal{N}$ of mass $m$ in the simulation following the analytical approximation in Eqn. \ref{eq:NNmdot}. 
 
\begin{figure*}
    \centering
    \makebox[\textwidth][c]{ 
        \begin{subfigure}[t]{0.37\textwidth}
            \centering
            \includegraphics[width=\linewidth]{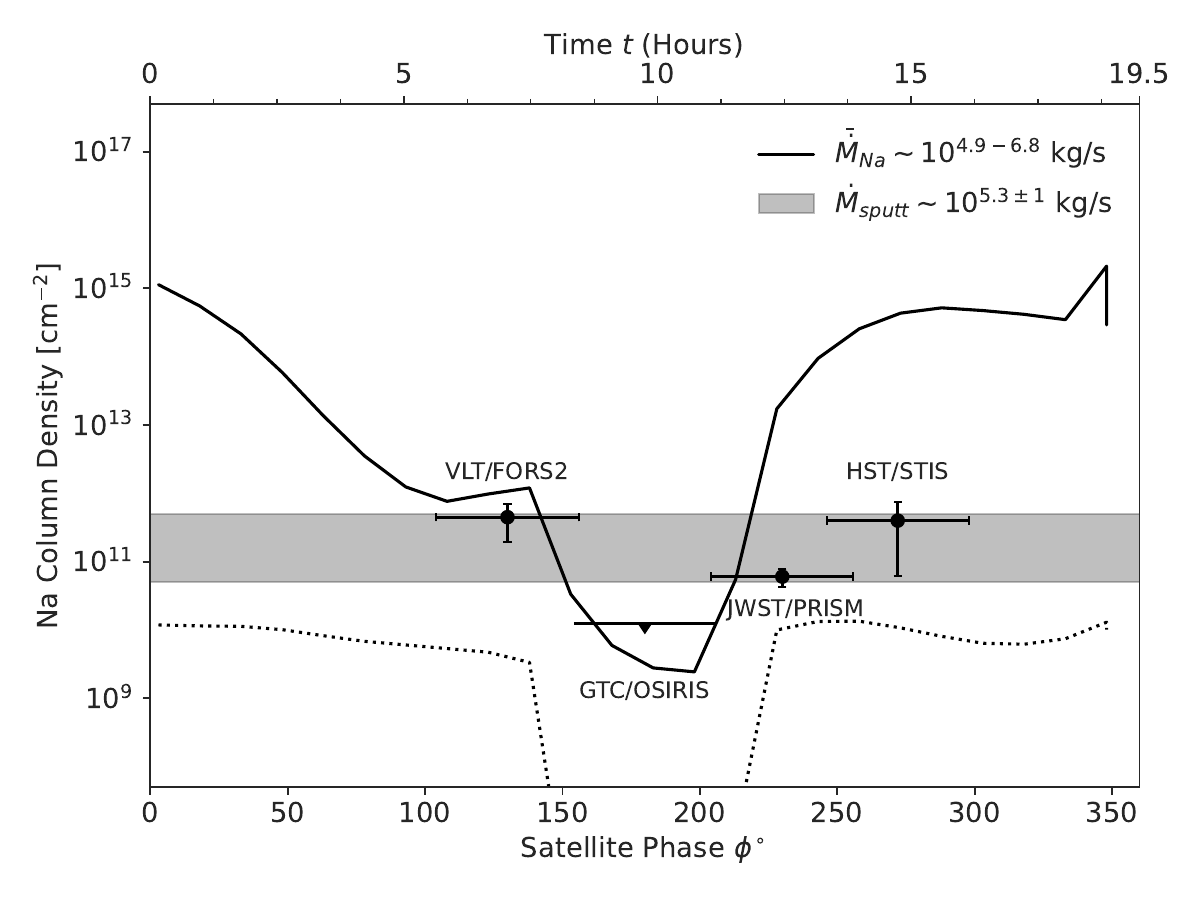}
            \caption{Neutral Sodium (Na I)}
            \label{fig:panel1}
        \end{subfigure}
        \hspace{-0.08em} 
        \begin{subfigure}[t]{0.37\textwidth}
            \centering
            \includegraphics[width=\linewidth]{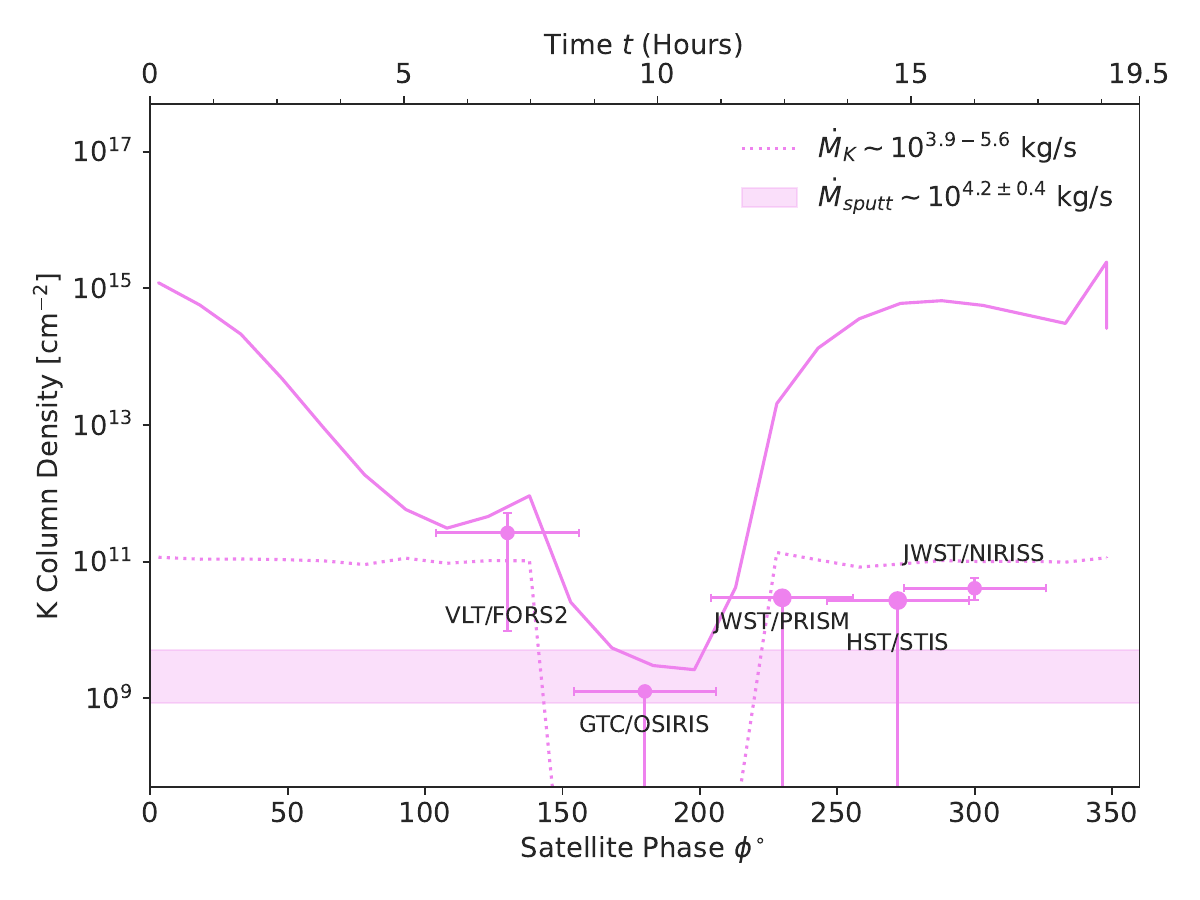}
            \caption{Neutral Potassium (K I)}
            \label{fig:panel2}
        \end{subfigure}
        \hspace{-0.08em}
        \begin{subfigure}[t]{0.37\textwidth}
            \centering
            \includegraphics[width=\linewidth]{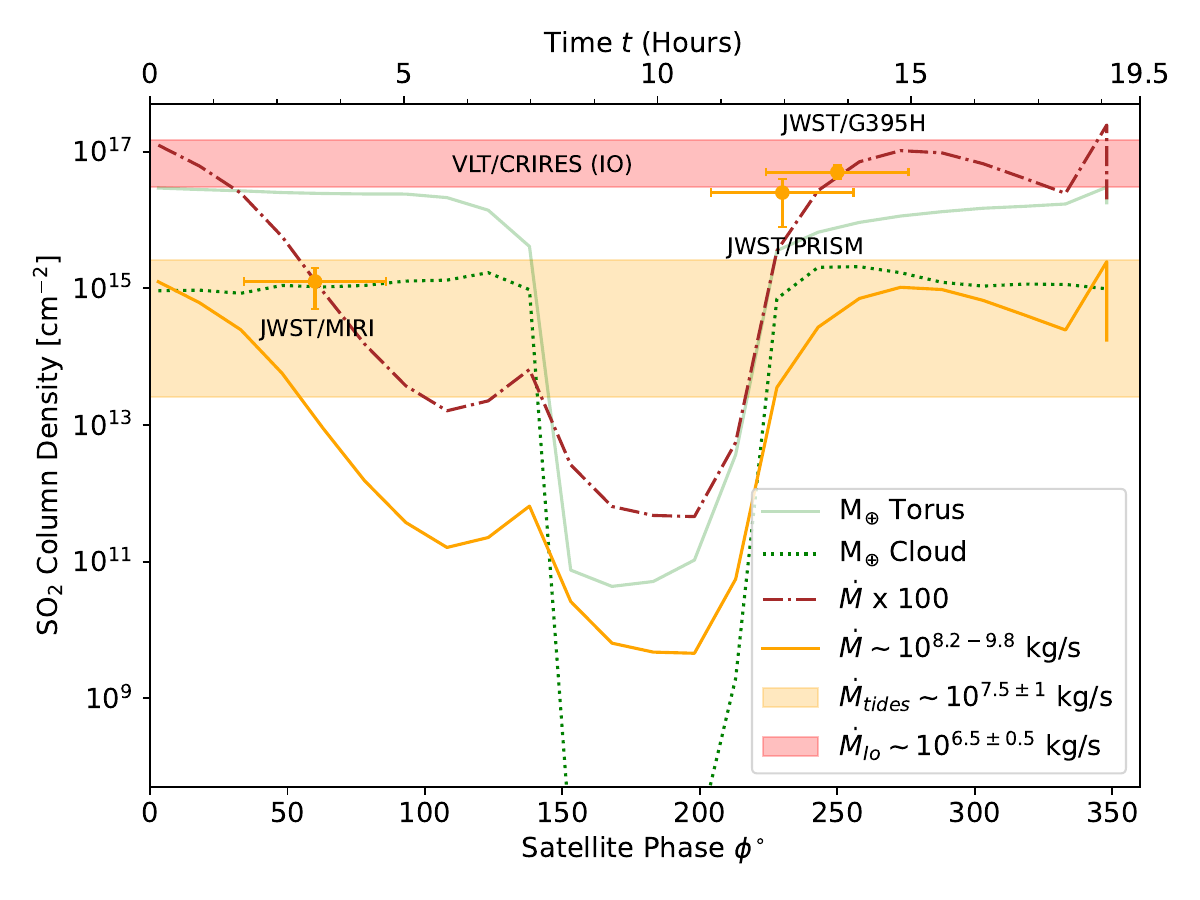}
            \caption{Neutral Sulfur Dioxide (SO$_2$)}
            \label{fig:panel3}
        \end{subfigure}
    }

    \caption{Minimum line-of-sight gas density variability of alkali metals and sulfur dioxide gas over several epochs.  Data points: measurement from each instrument/epoch for Na (a black), K (pink), and SO$_2$ (orange). X-errors: transit duration (2.8 hours). Horizontal bands: 1-D predictions from 3-body tidal heating estimates by \citet{oza2019b} for Na (gray), K (pink), and SO$_2$ (orange), with Io's observed column by VLT/CRIRES (red). Black, pink, orange lightcurves: 3-D Monte Carlo simulations \citep{MzW24} modelling sputtered clouds (dotted) and toroidal (solid) geometries. Panel iii: Green lightcurves: SO$_2$ simulations for an Earth-sized satellite cloud (dash-dot) and torus (solid). Brown dash-dot is $\sim$ 100$\times$ Io torus scenario in orange.}
    \label{fig:variability}
\end{figure*}

Considering the proximity of the star, we incorporate radiation pressure forces following the methodology outlined in \citet{MzW24}. We primarily consider R$_{\rightmoon} \sim R_{Io}$, although a terrestrial-sized volcanic satellite (e.g. \citet{BelloArufe2025})  is also considered in Figure \ref{fig:panel3}. Through analyzing the systems' evolution over time, we derive density \textit{phase-curves} that illustrate the average particle and line-of-sight densities at any point in time over the duration of one exomoon's orbit. Results for Na limited by photoionization at higher velocities yield an elongated cloud reminiscent of the Io Banana cloud \citep{SmythCombi1988, wilson02} (see Figure \ref{fig:serpensNaKSO2}). Albeit, the exomoon cloud is more localized and lacks a complete toroidal structure due to short particle lifetimes. The online version of this manuscript provides a time-series sequence of Na evolution for both tori (Fig. \ref{fig:serpens_left}) and cloud (Fig. \ref{fig:serpens_right}) scenarios. The variations in average particle densities $\bar{n}$ primarily stem from radiative forces, resulting in minor fluctuations compared to the changes in average column densities  $\bar{N}$. As lifetimes decrease further, particularly for K, evaporated material becomes increasingly confined to the exomoon, leading to a disappearance in line-of-sight signal during occultation. 
A prolonged species lifetime typically results in toroidal structures for small satellites. If the exomoon is massive the evaporated material may also return to the surface. These toroidal structures better distribute material along the orbit (similar to Io and Enceladus), which leads to less night-to-night variability while also increasing the total number of absorbing material. This is particularly interesting for SO$_2$, where column densities are large and more consistent between the data retrievals. Our torus simulations of SO$_2$ in Figure \ref{fig:serpensNaKSO2} suggest that column densities may vary by a significant amount but in a more controlled and continuous fashion. These trends are illustrated in Figure \ref{fig:panel1}, \ref{fig:panel2}, and \ref{fig:panel3} which show the temporal variability of integrated gas densities for Na, K, and SO$_2$ across different epochs and instruments as well as comparisons between simulated torus and cloud geometries under varying exomoon mass scenarios. We compute the maximum possible column density implied by a recent K I non-detection by Gran Telescopio Canarias (GTC)/OSIRIS \citep{Jiang2023} in Figure \ref{fig:panel2}. In contrast, a confined SO$_2$ cloud, strictly limited to the photoionization lifetime, can cause a complete loss of signal during exomoon occultation - more akin to something we expect for K. Furthermore, integrating the column density of a photoionization limited SO$_2$ cloud over the transit duration does not yield large enough values to explain what we see in data. Whether it is reasonable to assume a prolonged lifetime of SO$_2$ molecules at WASP-39 b is difficult to assess. Although we do not explicitly track ions, which are known to influence the Jupiter-Io plasma torus \citep{bagenal94}, we calculate photoionization and electron-impact. The ionization rates imply a large presence of free electrons and SO$_2$ ions, which may boost the lifetime on the order of a transit duration. 
\renewcommand{\thesubfigure}{\roman{subfigure}}

\vspace{-2em} 
\section{Conclusions}\label{sec:discussion}
Overall, we find that our numerical simulations in Figure \ref{fig:variability} are able to reproduce observed column densities, suggesting that a natural satellite source may not be negligible. For instance, especially for the Na and SO$_2$ cases in Figure \ref{fig:panel1} \& \ref{fig:panel3}, a torus is required in comparison to cloud densities (dotted line) are far too small no matter the mass of the unknown satellite. For Na an extended cloud , smaller than a torus, similar to the streams seen at Io approximates the epoch to epoch variability well. For K , since there is no identified KCl torus at Io, it may be possible that it is indeed a cloud, especially due to the disappearance of the K line completely in the JWST/PRISM data. Table \ref{tab:obs_table}'s mass loss ranges $\dot{M}$ indicate that a Na \& K torus is, in principle, able to be sustained based on JWST and VLT data of evolving absorption features from subtle alkali absorption (sputtered cloud geometry)  to large transit depths (torus geometry) . The $\dot{M}_{\gamma}$ - $\dot{M}_{tor}$ values provide a range required to sustain $N$. Characterizing possible instrumental and stellar variability would benefit from dedicated repeat observations.

\vspace{-1em} 
\subsection{Volcanic Exomoon Output in Time and Mass Flux}
\vspace{-0.8em} 
Results from the \texttt{serpens} simulations reveal that mass-loss from an exomoon can produce strongly localised Na \& K clouds. When integrating average Na column densities over the transit duration of $\sim$2.8 hours, we are able to infer values comparable in magnitude to those in \citet{oza2019b}. If K is locally bound to a putative exomoon, a natural explanation for a non-detection is the (partial) occultation of the satellite (and its exosphere) by the planet. 
Furthermore, as can be drawn from the phase-curve in Figure \ref{fig:variability}, a
toroidal structure for SO$_2$ (green solid line) is better at explaining the subtly varying observational column densities while also providing large enough values when integrating over the transit duration. Sputtered clouds imply \textit{orders} of magnitude variations, resulting in disappearances interpreted in Figure \ref{fig:panel1} and \ref{fig:panel2} as alkali clouds, in contrast to the volcanic gas tori implying a more subtle variation between epochs $\sim$ 20. The behavior is further consistent with \texttt{prometheus} geometries of an exomoon cloud and tori reported in Table \ref{tab:alkalis} and Figure \ref{fig0:NaK}.

\begin{figure}
    \centering
    \includegraphics[width=0.45\textwidth]{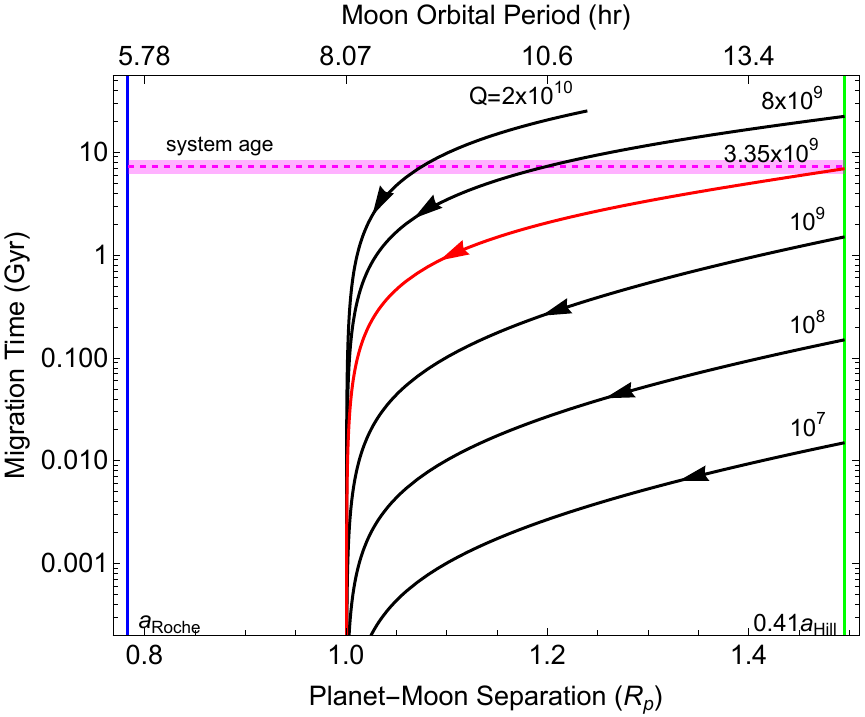}
        \caption{Moon migration plot for an exo-Io orbiting WASP-39~b. Each curve represents the time it takes for the moon to migrate due to equilibrium tides from a certain initial separation to the surface of the planet, for a given $\mathcal{Q}$ of the planet. The red curve pertains to the estimated $\mathcal{Q}$ of WASP-39~b based on the analysis in \citealt{oza2019b}. We assume the moon orbits faster than the planet spins, so that the moon falls towards the planet (towards the left of the plot). The pink region is the estimated age range of the system,  at 7.2~Gyr . The vertical lines represent the Roche limit (blue) and Hill sphere (green).}
    \label{fig:moonstable}
\end{figure}

The values inferred in Figure \ref{fig:variability} offer a range of possibilities for a putative exomoon to source the volcanic gases detected in transmission. By considering a cloud or toroidal lifetime $\tau_i$ as described in Section \ref{sec:alkali_analysis} following $\dot{M} \sim mN\pi R_\star ^2 /\tau_i$, we constrain roughly $\dot{M} \sim$ 10$^{4.9 - 6.8}$, 10$^{3.9 - 5.6}$, and 10$^{8.2-9.8}$ kg/s  for Na I (black), K I (pink), and SO$_2$ (orange) respectively, consistent with approximations predicted by \citet{oza2019b} (shaded regions) if a toroidal geometry is assumed \citep{johnson06b, MeyerZuWestram2023}. The green phase curve of an Earth-mass satellite, fits the data better than an Io-mass satellite (orange). Of course, it is also possible that a smaller satellite is interacting with the upper atmosphere of the gas giant,  similar to Enceladus and Io, as the escape would be far larger \citep{MzW24}. In that case, the mass loss ranges may imply the putative satellite is in its last stage of evolution, on its way to transform into a ring satellite. Due to the favorable orbit of WASP-39 b, it may be that this satellite is still intact (Figure \ref{fig:moonstable}), and has not yet fully disintegrated at the Roche limit (blue vertical line). The preferred tidal Q value from \citet{oza2019b}, red line in Figure \ref{fig:moonstable} implies a migration time $>$ 10 Gyrs to an orbit at the photosphere of $\sim$ 8.07h, assuring the survival of the satellite. The red band indicated in Figure \ref{fig:variability} is the column density of Io, with the mass loss range of volcanism needed to supply it against atmospheric sputtering \citep{johnson04}. Hot Saturns in addition to WASP-39 b have now been modeled in 3-D along with predictions of their NaD phase curves for HD209458 b, HD189733 b, WASP-49, WASP-96, WASP-69, WASP-17 b, XO-2N b, and HAT-P-1 b by \citet{MzW24}. 
\vspace{-2.6em}
\vspace*{0.1in}
\subsection{Outlook}

Toroidal exosphere simulations from \texttt{serpens} and \texttt{prometheus} approximate a line-of-sight (LOS) column $N \sim$ 10$^{16}$ SO$_2$/cm$^2$ from a tidally-heated outgassing rate predicted in previous modelling at $\sim$ 10$^{8.5}$ kg/s. Surprisingly this minimum rate for a toroidal geometry, changes by an order of magnitude for SO$_2$ as well, suggesting an exogenic process. The SO$_2$ column is nearly identical to Io's, whose disk is covered in infrared-bright features near 4 $\mu$m \citep{Carlson1997} and 7.3 $\mu$m as seen by ongoing \textit{Juno}/JIRAM \citep{Tosi2020} \&  JWST/MIRI observations \citep{deKleerMIRI2023}. Evaluating Na/SO$_2$, the neutral observations vary from $\sim$2500 to 10 $\times$ \textit{less} than Io's known $\chi_{NaCl/SO_2}$ abundance \citep{lellouch03}, likely due to enhanced escape and heating beyond 1100 K. Interestingly, the Roche limit for a rock is interior to the planetary surface, therefore atmosphere/exosphere interactions are especially important at this system. To constrain putative satellite orbits, high-resolution alkali Doppler shift observations are required \citep{Oza2024, Unni2025}. Since volcanic gas is expected to be accompanied by dust, recent JWST mid-infrared observations are further suggestive of a natural satellite at this alkali hot Saturn system \citep{Flagg2024} as well as other hot Jupiter systems studied in \citet{oza2019b} such as HD-189733 b \citep{Inglis2024} .

\subsection*{Acknowledgments}
A.V.O thanks H. Knutson for valuable insight on HST/STIS and N. Nikolov on VLT/FORS2 data. The research described in this paper was carried out in part at the Jet Propulsion Laboratory, California Institute of Technology, under a contract with the National Aeronautics Space Administration. © 2025. All rights reserved. Y.M. acknowledges support from the European Research Council (ERC) under the European Union’s Horizon 2020 research and innovation programme (grant agreement no. 101088557, N-GINE). 
\vspace{-2.5em} 
\section*{Data Availability}
Relevant JWST data used in this study can be accessed through the Space Telescope Science Institute (STScI) archive at https://archive.stsci.edu.

\bibliographystyle{mnras}
\bibliography{exo-io2023} 


\bsp	
\label{lastpage}
\end{document}